\documentclass[aps,prl,twocolumn,preprintnumbers,superscriptaddress]{revtex4-1}

\usepackage[usenames,dvipsnames,table]{xcolor}
\usepackage{graphicx,amsmath,amssymb,amsthm,multirow,array,bm,soul}
\usepackage[mathscr]{eucal}
\usepackage[bbgreekl]{mathbbol}
\usepackage{amsfonts}
\usepackage{hyperref}
\usepackage{tikz}
\usetikzlibrary{3d,calc}

\newcommand{\beq}{\begin{equation}}
\newcommand{\eeq}{\end{equation}}
\newcommand{\II}{\mathrm{I}\hspace{-0.8pt}\mathrm{I}}
\def\<{\langle}
\def\>{\rangle}

\newcommand{\cuv}{c_{\textrm{UV}}}
\newcommand{\cir}{c_{\textrm{IR}}}
\newcommand{\see}{S_{\textrm{EE}}}

\newcommand{\nn}{\nonumber}
\newcommand{\sdef}{S^{\textrm{def}}_{\textrm{EE}}}

\newcommand{\N}{{\mathcal{N}}}

\begin{document}

\title{From the Weyl Anomaly to Entropy of Two-Dimensional Boundaries and Defects}

\author{Kristan Jensen}
\email{kristanj@sfsu.edu}
\affiliation{Department of Physics and Astronomy, San Francisco State University, San Francisco, CA 94132, USA}

\author{Andy O'Bannon}
\email{a.obannon@soton.ac.uk}
\affiliation{STAG Research Centre, University of Southampton, Highfield, Southampton SO17 1BJ, UK}

\author{Brandon Robinson}
\email{B.J.Robinson@soton.ac.uk}
\affiliation{STAG Research Centre, University of Southampton, Highfield, Southampton SO17 1BJ, UK}

\author{Ronnie Rodgers}
\email{R.J.Rodgers@soton.ac.uk}
\affiliation{STAG Research Centre, University of Southampton, Highfield, Southampton SO17 1BJ, UK}

\date{\today}

\begin{abstract}
We study whether the relations between the Weyl anomaly, entanglement entropy (EE), and thermal entropy of a two-dimensional (2D) conformal field theory (CFT) extend to 2D boundaries of 3D CFTs, or 2D defects of $D \geq 3$ CFTs. The Weyl anomaly of a 2D boundary or defect defines two or three central charges, respectively. One of these, $b$, obeys a c-theorem, as in 2D CFT. For a 2D defect, we show that another, $d_2$, interpreted as the defect's ``conformal dimension,'' must be non-negative if the Averaged Null Energy Condition (ANEC) holds in the presence of the defect. We show that the EE of a sphere centered on a planar defect has a logarithmic contribution from the defect fixed by $b$ and $d_2$. Using this and known holographic results, we compute $b$ and $d_2$ for 1/2-BPS surface operators in the maximally supersymmetric (SUSY) 4D and 6D CFTs. The results are consistent with $b$'s c-theorem. Via free field and holographic examples we show that no universal ``Cardy formula'' relates the central charges to thermal entropy.
\end{abstract}

\maketitle

\textit{Introduction.}~CFTs play a central role in many branches of physics. In condensed matter physics they describe critical points. In string theory the worldsheet theory is a CFT. In quantum field theory (QFT) CFTs are fixed points of renormalization group (RG) flows.

CFTs in two-dimensional Minkowski space, i.e. 2D CFTs, enjoy Virasoro symmetry with central charge $c$. Unitarity plus ground state normalizability imply $c \geq 0$. For an RG flow from ultra-violet (UV) to infra-red (IR) CFTs with central charges $\cuv$ and $\cir$, respectively, unitarity, locality, and Poincar\'e symmetry imply the c-theorem: $\cuv \geq \cir$~\cite{Zamolodchikov:1986gt}. These properties suggest $c$ measures the effective number of massless degrees of freedom (DOF), which we indeed expect to be non-negative and to decrease along RG flows, as massive modes decouple.

Virasoro symmetry also implies that $c$ determines at least four other quantities that can count DOF in other ways. First, $c$ fixes the normalization of the 2-point function of the stress-energy tensor, $T^{\mu\nu}$. Second, conformal symmetry requires $T^{\mu}_{~\mu}=0$, but with a non-trivial spacetime metric $g_{\mu\nu}$ with Ricci scalar ${\cal R}$, quantum effects can produce the Weyl anomaly, $T^{\mu}_{~\mu}=-\frac{c}{24\pi} {\cal R}~$\cite{Wess:1971yu,Bonora:1985cq,Deser:1993yx,Duff:1993wm}. Third, in a CFT's vacuum the EE of a spatial interval of length $\ell$, which measures the strength of vacuum correlations, is $\see= \frac{c}{3}\ln(\ell/\varepsilon)+{\cal O}(\varepsilon^0)$~\cite{Holzhey:1994we,Calabrese:2004eu}, with UV cutoff $0<\varepsilon\ll1$. Fourth, at non-zero temperature $T$, Cardy showed that the CFT's entropy density, $s$, which measures the number of thermodynamic microstates, is $s = \frac{\pi}{6} \, c \,T$~\cite{Cardy:1986ie,Affleck:1986bv}. 

CFTs have an infinite correlation length. However, no real system is infinite: boundary conditions (BCs) will always be important. Moreover, no real system is perfect: defects such as impurities, domain walls between differently ordered phases, and so on will always be important. Constructing and classifying CFTs with conformally-invariant boundaries (BCFTs) or defects (DCFTs) is thus crucial for describing an enormous number of systems, including impurities in metals, graphene, Yang-Mills (YM) Wilson and 't Hooft lines, D-branes, and more.

In this Letter we study a 2D boundary of a 3D CFT or 2D conformal defect in a $D\geq 3$ CFT. We assume the boundary or defect is flat, i.e. a static straight line. At least two arguments show that such a system does not have Virasoro symmetry in general. First, the 2D contribution to $T^{\mu\nu}$ is not conserved because energy and momentum can flow between the boundary or defect and the CFT. Second, the Virasoro algebra has an infinite number of generators, but these systems have a finite number: the CFT's $SO(D,2)$ conformal symmetry is broken to the subgroup that leaves the boundary or defect invariant, $SO(2,2) \times SO(D-2)$, where $SO(2,2)$ are conformal transformations leaving the static line invariant and $SO(D-2)$ are rotations about the static line.~\footnote{Every independently conserved spin-2 operator at the boundary or defect may give rise to a Virasoro symmetry. However in general such operators appear only if the 2D degrees of freedom decouple from the bulk CFT, or if the bulk CFT has no propagating degrees of freedom. An example of the latter is 3D Chern-Simons theory, which is topological and hence has vanishing stress tensor.}

We will address the natural questions that arise for 2D defects or boundaries in the absence of Virasoro symmetry: what do $T^{\mu}_{~\mu}$, $\see$, and $s$ look like? Are they still related in the same way as in a 2D CFT?

We find that certain EE's have a logarithmic term fixed by $T^{\mu}_{~\mu}$, while in general no simple relation exists between $T^{\mu}_{~\mu}$ and $s$. The boundary or defect contribution to $T^{\mu}_{~\mu}$ includes not only ${\cal R}$ but also extrinsic curvature contributions, leading to multiple central charges~\cite{Henningson:1999xi,Schwimmer:2008yh}. Assuming the ANEC holds in the presence of the defect, we place a new bound on one of these. (In Appendix A we also find new central charges allowed in 4D if parity is broken.) For $\see$ of a spherical region centered on a defect, we use the method of refs.~\cite{Casini:2011kv,Jensen:2013lxa}, involving a conformal transformation from Minkowski to de Sitter space, to show that $\see$'s term $\propto \ln(\ell/\varepsilon)$ in general depends on \textit{two} defect central charges. (This result agrees with and extends a key result of ref.~\cite{Kobayashi:2018lil}.) Using this and known holographic results, we compute these central charges for certain 1/2-BPS surface operators in the maximally SUSY 4D and 6D CFTs. Finally, we calculate $s$ for the free, massless scalar and fermion 3D BCFTs and for 2D defects holographically dual to probe branes. We find a term $\propto T$ at the boundary or defect, but show that any putative relation between its coefficient and the central charges cannot be universal.
 

\textit{The Systems.}~We start with a local, unitary, Lorentzian CFT on a $D\geq 3$ spacetime $\cal{M}$ with coordinates $x^{\mu}$ ($\mu=0,1,\ldots,D-1$) and metric $g_{\mu\nu}$, which we will call the ``bulk'' CFT. We introduce a codimension $D-2$ 
defect along a static 2D submanifold $\Sigma$ with coordinates $y^a$ ($a = 0,1$). We parameterize $\Sigma\hookrightarrow \cal{M}$ by embedding functions $X^{\mu}(y)$ such that $\Sigma$'s induced metric is $\gamma_{ab} \equiv \partial_a X^{\mu} \partial_b X^{\nu} \, g_{\mu\nu}$. We denote $\cal{M}$'s covariant derivative as $\nabla_\mu$ and $\Sigma$'s induced covariant derivative as $\hat{\nabla}_a$, which acts on a mixed tensor ${\cal T}^{\mu}_{~~a}$ as $\hat{\nabla}_a{\cal T}^{\mu}_{~~b} = \partial_a {\cal T}^{\mu}_{~~b} + \Gamma^{\mu~}_{\nu a} {\cal T}^{\nu}_{~~b} -\hat{\Gamma}^c_{ab}{\cal T}^{\mu}_{~~c}$. The second fundamental form is then $\Pi^{\mu}_{~ab} = \hat{\nabla}_a \partial_b X^{\mu}$, with traceless part $\mathring{\Pi}^{\mu}_{~ab} \equiv \Pi^{\mu}_{~ab} - \frac{1}{2}\gamma_{ab} \gamma^{cd}\Pi^{\mu}_{~cd}$.

Physically, the defect can arise from 2D DOF coupled to the bulk CFT and/or BCs imposed on bulk CFT fields~\footnote{In the replica trick approach to calculating R\'{e}nyi entropies in a 4D CFT, twist operators appear as 2D conformal defects defined by boundary conditions on bulk CFT fields. Those defect's central charges can then determine both universal~\cite{Lewkowycz:2014jia,Perlmutter:2015vma,Bianchi:2015liz} and non-universal~\cite{Ohmori:2014eia} terms in EE.}. In 3D the defect is a domain wall between two CFTs, and if one of these is the ``trivial'' CFT, then the defect is a boundary. Our results will thus apply to 3D BCFTs, but for brevity we will only explicitly discuss DCFTs unless stated otherwise.
 

\textit{The Weyl Anomaly.}~We denote the DCFT partition function as $Z$. The generating functional of connected correlators is then $W \equiv - i \ln Z$. Both are functionals of $g_{\mu\nu}$ and $X^{\mu}$. We define the stress-energy tensor, $T_{\mu\nu}$, and displacement operator, $\cal{D}_{\mu}$, by variations of $W$,
\beq
\delta W= \frac{1}{2}\int d^Dx  \sqrt{-g}\, \delta g_{\mu\nu}\langle T^{\mu\nu} \rangle +\int d^2y \sqrt{-\gamma} \,\delta X^{\mu} \langle {\cal{D}}_{\mu}\rangle, \nn
\eeq
where $g\equiv\det g_{\mu\nu}$ and $\gamma\equiv\det\gamma_{ab}$. Invariance of $W$ under reparametrizations of $y^a$ implies $\cal{D}_{\mu}$'s components along $\Sigma$ vanish~\cite{Jensen:2015swa}. Invariance of $W$ under reparametrizations of $x^{\mu}$ implies $\nabla_{\nu}\langle T^{\nu\mu} \rangle =  - \delta^{D-2} \langle {\cal D}^{\mu}\rangle$, with $\delta^{D-2}$ a delta function that restricts to $\Sigma$~\cite{Jensen:2015swa}. Physically $\langle T^{\mu\nu}\rangle$ is not conserved at $\Sigma$ because the defect and bulk can exchange transverse energy-momentum.

Our DCFTs are invariant under infinitesimal Weyl transformations, $\delta_\omega g_{\mu\nu} = 2\omega g_{\mu\nu}$ and $\delta_\omega X^\mu =0$, only up to the Weyl anomaly~\cite{Wess:1971yu,Bonora:1985cq,Deser:1993yx,Duff:1993wm}: $\delta_{\omega} W =\int d^Dx \sqrt{-g}\, \omega \langle T^{\mu}_{~\mu} \rangle$, where $\langle T^{\mu}_{~\mu} \rangle$ is built from external sources, such as $g_{\mu\nu}$. We will consider contributions to $\langle T^{\mu}_{~\mu} \rangle$ from $g_{\mu\nu}$ and $\partial_a X^{\mu}$ only. Determining $\langle T^{\mu}_{~\mu} \rangle$'s most general form requires solving the Wess-Zumino (WZ) consistency condition, which comes from the fact that two successive Weyl transformations of $W$ commute, and then fixing any local counterterms that contribute to $\langle T^{\mu}_{~\mu} \rangle$. In our DCFTs, $\langle T^{\mu}_{~\mu} \rangle = \langle T^{\mu}_{~\mu} \rangle_{\textrm{bulk}} +\delta^{D-2}\langle T^{\mu}_{~\mu} \rangle_{\textrm{def}}$, where $\langle T^{\mu}_{~\mu} \rangle_{\textrm{bulk}}$ and $\langle T^{\mu}_{~\mu} \rangle_{\textrm{def}}$ are bulk CFT and defect Weyl anomalies, respectively, and we fixed local counterterms to cancel terms with normal derivatives of $\delta^{D-2}$. For $\langle T^{\mu}_{~\mu} \rangle_{\textrm{bulk}}$ we will only need to know that $\langle T^{\mu}_{~\mu} \rangle_{\textrm{bulk}}=0$ in odd $D$, but can be non-zero in even $D$, which defines the bulk central charge(s). For a 2D defect in a $D\geq 3$ DCFT~\cite{Graham:1999pm,Henningson:1999xi,Schwimmer:2008yh},
\beq
\label{eq:defect-A}
\langle T^{\mu}_{~\mu} \rangle_{\textrm{def}} =-\frac{1}{24\pi}\left( b \, {\cal{R}}_\Sigma + d_1 \, \mathring{\II}^{\mu}_{ab}\mathring{\II}_{\mu}^{ab} - d_2 \, W_{ab}^{~~ab} \right),
\eeq
where ${\cal{R}}_\Sigma$ is $\Sigma$'s intrinsic scalar curvature, $W_{abcd}$ is the pullback of the bulk Weyl tensor to $\Sigma$, and $b$, $d_1$, and $d_2$ are defect central charges. When $D=3$, $W_{abcd}=0$ identically, so $d_2$ exists only for $D \geq 4$.

\textit{Bounds on Central Charges.} As mentioned above, in a 2D CFT $c$ determines various observables and with reasonable assumptions, such as unitarity, obeys $c \geq 0$ and the c-theorem. By comparison, less is known about $b$, $d_1$, and $d_2$. Under Weyl transformations $\sqrt{-\gamma}\,{\cal R}_{\Sigma}$ changes by a total derivative (type A in the classification of ref.~\cite{Deser:1993yx}), while both $\sqrt{-\gamma}\mathring{\II}^{\mu}_{ab}\mathring{\II}_{\mu}^{ab}$ and $\sqrt{-\gamma}W_{ab}^{~~ab}$ are invariant (type B). As a result, in the Euclidean DCFT on $\mathbb{S}^D$ of radius $r$ with bulk partition function $Z_{\textrm{CFT}}$ and with defect along a maximal $\mathbb{S}^2$, $Z/Z_{\textrm{CFT}}\propto \left(r\Lambda\right)^{b/3}$~\cite{Jensen:2015swa}. For a local, unitary defect RG flow, $b$ obeys a c-theorem, suggesting $b$ counts defect DOF~\cite{Jensen:2015swa}. WZ consistency forces $b$ to be independent of any marginal couplings. The normalization of ${\cal D}^{\mu}$'s 2-point function is fixed by $d_1$, such that unitarity implies $d_1 \geq 0$~\cite{Herzog:2017xha,Herzog:2017kkj}.

\begin{table}
\centering
\begin{tabular}{| c | c | c | c |c|}
\hline
Theory & BC & $b$ & $d_1$ & $d_2$ \\\hline
Scalar & Dirichlet & $-1/16$ & $3/32$ & N/A \\\hline
Scalar & Robin & $1/16$ & $3/32$ & N/A \\\hline
Fermion & Mixed & $0$ & $3/16$ & N/A \\\hline
Probe Brane & N/A & $6\pi L^3T_{\textrm{br}}$ & $6 \pi L^3T_{\textrm{br}}$ & $6\pi L^3T_{\textrm{br}}$ \\
\hline
\end{tabular}
\caption{\label{tab:ccs}Central charges $b$, $d_1$, and $d_2$ of eq.~\eqref{eq:defect-A} for 3D BCFTs of free, massless real scalars with Dirichlet or Robin BC, or Dirac fermion with the unique conformal ``mixed'' BC~\cite{Jensen:2015swa,Nozaki:2012qd,Fursaev:2016inw}, and for a 2D defect dual to a probe brane of tension $T_{\textrm{br}}$ along $AdS_3$ inside $AdS_{D+1}$ of radius $L$~\cite{Graham:1999pm}.}
\end{table}

Table~\ref{tab:ccs} shows $b$, $d_1$, and $d_2$ in the 3D BCFTs of a free, massless real scalar or Dirac fermion~\cite{Jensen:2015swa,Nozaki:2012qd,Fursaev:2016inw}, and in CFTs holographically dual to Einstein gravity in $(D+1)$-dimensional Anti-de Sitter space, AdS$_{D+1}$, with metric $G_{MN}$ ($M,N=0,1,\ldots,D$) and defect dual to a probe brane along AdS$_3$ whose action $S_{\textrm{probe}} = - T_{\textrm{br}} \int d^3\xi \sqrt{- \text{det}(P[G_{MN}])}$ with tension $T_{\textrm{br}}$ and brane coordinates $\xi$~\cite{Graham:1999pm}. In all of these theories, the central charges are $\geq 0$ with the exception of the scalar with Dirichlet BC, which has $b < 0$. This example proves that unitarity does not require $b \geq 0$.

Indeed, for a unitary 3D BCFT with unique stress-energy tensor at the boundary~\footnote{Every independently conserved spin-2 operator at the boundary, with associated central charge $c$, contributes a term $\propto c$ to $b$ in addition to those in eq.~\eqref{E:bfromTT}.}, ref.~\cite{Herzog:2017kkj} conjectured
\beq
\label{E:bfromTT}
b = \frac{2\pi^2}{3}\,\epsilon(1) - \frac{2}{3}\,d_1,
\eeq
where $\epsilon(v)$ is a contribution to $T^{\mu\nu}$'s 2-point function from exchange of spin-2 boundary operators, with $v\in[0,1]$ the BCFT conformal cross ratio, with boundary at $v=1$~\cite{McAvity:1993ue,McAvity:1995zd,Herzog:2017xha}. Unitarity implies $\epsilon(v) \geq 0$~\cite{Herzog:2017xha}. However, if the BCFT has any spin-2 boundary operators of dimension $\Delta \in [2,3)$, then $\epsilon(v)$ diverges as $(1-v)^{\Delta - 3}$ when $v \to 1$. In that case, unitarity does not constrain the sign of $\epsilon(1)$, the $(1-v)^0$ term in $\epsilon(v)$'s expansion about $v=1$, and so $b$ has no definite lower bound. On the other hand, in the absence of such operators $\epsilon(v)$ is regular as $v \to 1$, unitarity implies $\epsilon(1) \geq 0$, and hence $b \geq - \frac{2}{3}d_1$. All the examples in table.~\ref{tab:ccs} obey this bound, and the free scalar with Dirichlet BC saturates it.

We can prove a new bound, $d_2\geq0$, assuming the ANEC holds in the presence of the defect. The ANEC states that for any null direction $u$, $\int_{-\infty}^\infty du \langle T_{uu}\rangle \geq 0$, meaning the total energy measured by the light-like observer along $u$ is $\geq 0$. Proofs of the ANEC for CFTs appear in refs.~\cite{Faulkner:2016mzt,Hartman:2016lgu}. Though these proofs have not yet been extended to BCFTs or DCFTs, they rely mainly on unitarity and causality, which in a BCFT or DCFT should suffice to guarantee that a light-like observer's total energy is $\geq 0$.

\begin{figure}[t]
\begin{center}
\includegraphics[width = 0.875\columnwidth]{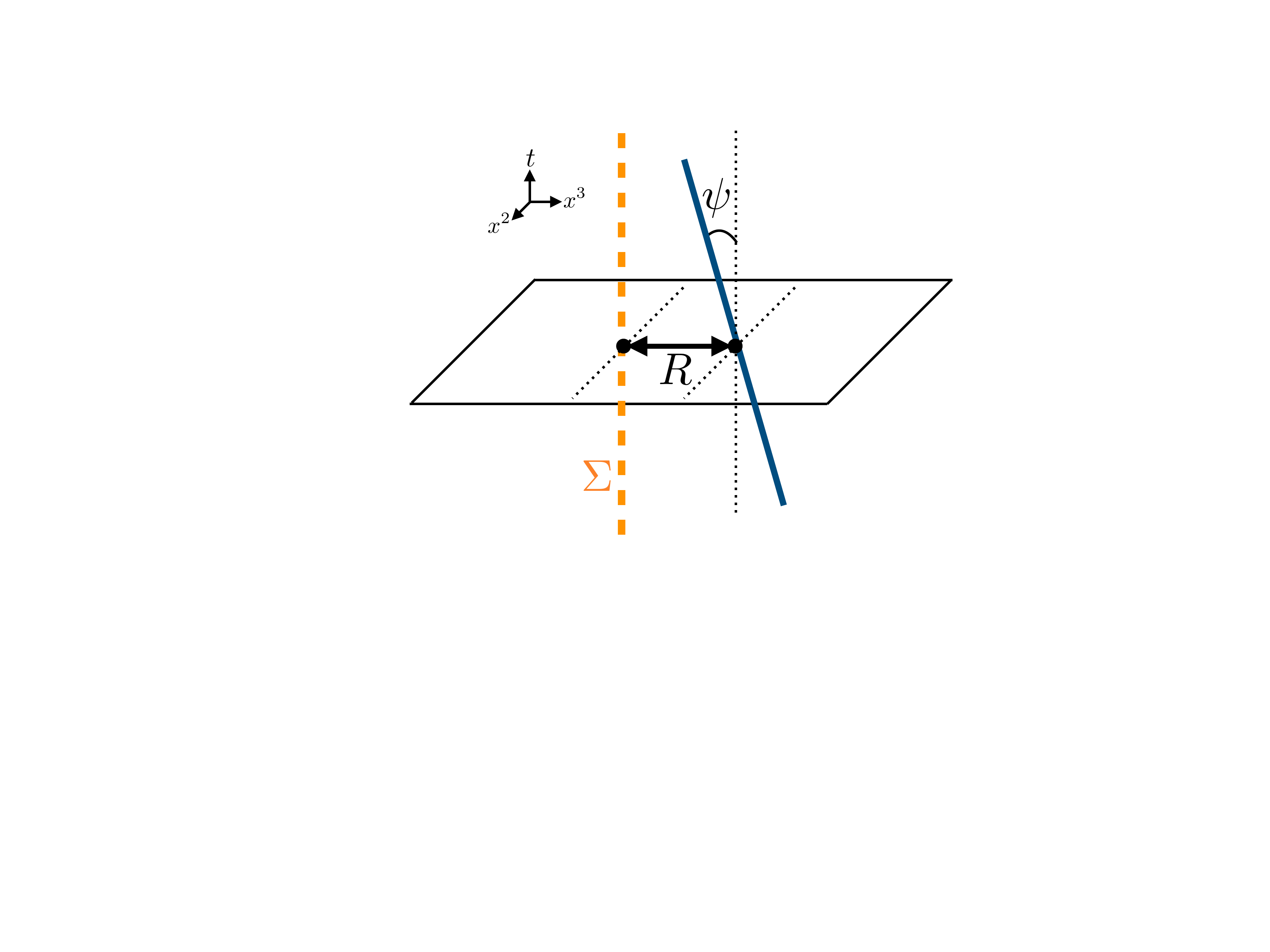}
\caption{The orange dashed line is the 2D defect along $\Sigma$. The solid blue line is the null geodesic in eq.~\eqref{eq:geo}, at a distance $R$ and angle $\psi$ from $\Sigma$.
}\label{fig:anec-fig}. \end{center}
\end{figure}

For a CFT in Minkowski space, $SO(D,2)$ symmetry forces $\langle T_{\mu\nu}\rangle=0$. However, in our DCFTs when $\Sigma$ has co-dimension $\geq 2$, $SO(2,2) \times SO(D-2)$ symmetry allows $\langle T_{\mu\nu} \rangle \neq 0$. In fact, when $D=4$ refs.~\cite{Lewkowycz:2014jia,Bianchi:2015liz} showed that $\langle T_{\mu\nu} \rangle$ is completely determined by $d_2$ by writing the most general form of $\langle T_{\mu\nu} \rangle$ allowed by DCFT symmetry, using differential regularization to make $\langle T_{\mu\nu} \rangle$ well-defined as a distribution, and comparing its variation under constant Weyl transformations to the variation of $\langle T^{\mu}_{~\mu} \rangle_{\textrm{def}}$ with respect to $g_{\mu\nu}$. Generalizing the result of refs.~\cite{Lewkowycz:2014jia,Bianchi:2015liz} to any $D$ is straightforward: with coordinates $x^i$ transverse to $\Sigma$ ($i = 2, 3, \ldots, D-1$), and for a point a distance $|x^i|>0$ from $\Sigma$,
\begin{align}
\label{eq:T1pt}
\begin{split}
\langle T^{ab}\rangle & = -\frac{h_D}{2\pi}\frac{\eta^{ab}}{|x^i|^D}\,, \qquad \langle T^{ai}\rangle = 0\,,
\\
\langle T^{ij}\rangle & = \frac{h_D}{2\pi(D-3)}\frac{3\delta^{ij} |x^k|^2 - D x^i x^j}{|x^l|^{D+2}}\,,\\
& h_D \equiv \frac{1}{3\text{vol}(\mathbb{S}^{D-3})} \, \frac{D-3}{D-1} \, d_2,
\end{split}
\end{align}
where $h_D$ is the defect's ``conformal dimension'' (see e.g. ref.~\cite{Kapustin:2005py}). Using $SO(2,2)\times SO(D-2)$  transformations, any null geodesic a distance $R$ from $\Sigma$ can be mapped to
\beq
\label{eq:geo}
t = Ru, \,\,\, x^1 = Ru \cos\psi, \,\,\, x^2=Ru \sin\psi, \,\,\, x^3 =R,
\eeq
and $x^{i>3}=0$, where $\psi$ is the angle between $\Sigma$ and the null geodesic, as shown in fig.~\ref{fig:anec-fig}. 
Plugging eqs.~\eqref{eq:T1pt} and~\eqref{eq:geo} into the ANEC gives
\begin{align}
\int_{-\infty}^{\infty}du\langle T_{uu}\rangle = \frac{1}{6\sqrt{\pi}R^D}\frac{|\sin\psi|}{\textrm{vol}(\mathbb{S}^{D-3})}\frac{\Gamma(\frac{D-1}{2})}{\Gamma(\frac{D}{2})}\,d_2\geq 0,
\end{align}
which immediately implies $d_2 \geq 0$.

\textit{EE and Central Charges.} Consider a CFT in $D$-dimensional Minkowski space with a flat 2D defect. We will compute the EE of a sphere of radius $\ell$ centered on the defect, using the method of refs.~\cite{Casini:2011kv,Jensen:2013lxa}.

We parameterize the Minkowski metric as
\beq
\eta = -dt^2 + (dx^1)^2 + (d|x^i|)^2 + |x^i|^2 \, ds_{\mathbb{S}^{D-3}}^2,
\eeq
with the defect along $t$ and $x^1$ and located at $|x^i| = 0$. Defining $r^2 = (x^1)^2 + |x^i|^2$, the sphere's causal development is given by $r \pm t \leq \ell$. The change of coordinates
\begin{align}
\begin{split}
	t   = \frac{\ell\cos\theta \sinh\left(\frac{\tau}{\ell}\right)}{1+\cos\theta\cosh\left( \frac{\tau}{\ell}\right)},& \quad r = \frac{\ell \sin\theta}{1 + \cos\theta \cosh\left(\frac{\tau}{\ell}\right)},\\
	x^1 = r \cos\phi,& \qquad |x^i| = r \sin\phi\,.
\end{split}
\end{align}
maps the sphere's causal development to the static patch of $D$-dimensional de Sitter space, $dS_{D}$, with metric
\begin{align}
\label{E:dS}
\Omega^2 \eta =-\ell^{-2}\cos^2\theta &d\tau^2 + d\theta^2 + \sin^2\theta \left( d\phi^2 + \sin^2\phi ds^2_{\mathbb{S}^{D-3}}\right),\nn\\
\Omega & = 1 + \cos\theta \cosh\left( \tau/\ell\right),
\end{align}
with $\tau \in (-\infty,\infty)$, $\theta \in [0,\pi/2]$, and $\phi \in [0,\pi]$. The defect is then along $\tau$ and $\theta$, and is located at $\phi = 0,\pi$, i.e. along a maximal dS$_2$.

The reduced density matrix of the sphere's causal development maps to $e^{-\beta H_{\tau}}$ modulo normalization, with $\beta = 2 \pi \ell$ and $H_{\tau}$ the generator of $\tau$ translations. As a result, $\see$ maps to thermal entropy in dS$_D$ at inverse temperature $\beta = 2 \pi \ell$,
\beq
\label{eq:dsent}
\see = \beta E - F,
\eeq
with $E$ the Killing energy corresponding to $H_{\tau}$ and $F = -\ln \textrm{tr} \left(e^{-\beta H_{\tau}}\right)$ the dimensionless free energy. We define the defect's contribution as $\see^{\textrm{def}} \equiv \see - \see^{\textrm{CFT}}$, with $\see^{\textrm{CFT}}$ the EE of a sphere of radius $\ell$ in the bulk CFT, with the same UV cutoff, and similarly for $E^{\textrm{def}}$ and $F^{\textrm{def}}$.

Analytically continuing to Euclidean time, $\tau = - i\tau_E$ with $\tau_E \sim \tau_E + \beta$, eq.~\eqref{E:dS} becomes the metric of $\mathbb{S}^D$, with the defect wrapping a maximal $\mathbb{S}^2$. As a result, $F= - \ln Z$, with $Z$ the DCFT's Euclidean partition function on $\mathbb{S}^D$. Using $Z/Z_{\textrm{CFT}}\propto \left(r\Lambda\right)^{b/3}$ with $\Lambda =1/\varepsilon$, we find
\beq
\label{eq:dsfree}
F^{\textrm{def}} = - \ln \left(Z/Z_{\textrm{CFT}}\right) = -\frac{b}{3}\ln \left( \frac{\ell}{\varepsilon}\right)+ {\cal O}(\varepsilon^0).
\eeq

The defect's contribution to the Killing energy is
\beq
E^{\textrm{def}} =  \int dS_{\mu}K_{\nu} \langle T^{\mu\nu}\rangle_{\textrm{def}},
\eeq
with $dS_{\mu}=\ell^{-1} \cos \theta (\sin\theta)^{D-2}(\sin\phi)^{D-3} d\theta d\phi\,ds_{\mathbb{S}_{D-3}}\, \delta_{\mu}^{\tau}$ the volume element on a constant time slice, and $K^{\mu}\partial_{\mu} = \partial_{\tau}$ the time translation Killing vector. We thus need $\langle T^{\tau}{}_{\tau} \rangle_{\textrm{def}}$, which we obtain by Weyl transformation from the Minkowski-space $\langle T^{\mu\nu}\rangle$ in eq.~\eqref{eq:T1pt}, with the result
\beq
\langle T^{\tau}{}_{\tau} \rangle_{\textrm{def}} = -\frac{(\sin\theta \sin\phi)^{-D}}{6\pi \text{vol}(\mathbb{S}^{D-3})}\frac{D-3}{D-1}d_2  - \frac{b}{24\pi}\delta^{D-2},
\eeq
where the first term comes from Weyl-rescaling $T^{\mu\nu}$ and the second term comes from $T^{\mu\nu}$'s anomalous Weyl transformation law. The integral of the second term is finite, but the integral of the first diverges at $\theta=0$ and the defect $\phi=0,\pi$. Using a regulator $\varepsilon/\ell$, we thus find
\beq
\label{eq:dsener}
\beta E^{\textrm{def}} = - \frac{1}{3}\frac{D-3}{D-1} \, d_2 \ln \left( \frac{\ell}{\varepsilon}\right) + {\cal O}(\varepsilon^0).
\eeq
Plugging eqs.~\eqref{eq:dsfree} and~\eqref{eq:dsener} into eq.~\eqref{eq:dsent} then gives~\footnote{These arguments can be generalized to the EE of a hemisphere of radius $r$ ending on the boundary of a 4D BCFT. We find that the boundary's contribution is equal to minus Gaiotto's ``boundary ${\cal F}$''~\cite{Gaiotto:2014gha},
\beq
 S_{\rm EE}^{\partial} \equiv S_{\rm EE} - \frac{1}{2}S_{\rm EE}^{\textrm{CFT}} = \ln \left(Z/\sqrt{Z_{\textrm{CFT}}}\right) \equiv - {\cal F}.
\eeq
Evidence from holography~\cite{Estes:2014hka} and free fields~\cite{Gaiotto:2014gha} suggests that $\mathcal{F}$ obeys a boundary c-theorem.}
\beq
\label{eq:eesphere}
\sdef = \frac{1}{3}\left( b - \frac{D-3}{D-1}d_2\right)\ln \left( \frac{\ell}{\varepsilon}\right) + {\cal O}(\varepsilon^0).
\eeq

As a check, we have also derived eq.~\eqref{eq:eesphere} via the replica method, generalizing the result of ref.~\cite{Fursaev:2016inw} for 3D BCFTs to $D \geq 3$ DCFTs.~\footnote{Ref.~\cite{Fursaev:2016inw} also considered EE of regions that do not intersect the boundary perpendicularly, finding a logarithmic term whose coefficient is a linear combination of $b$ and $d_1$. However, in such cases the replica symmetry is only $\mathbb{Z}_n$ with replica index $n$, in contrast to a CFT or a region that hits a boundary or defect perpendicularly, where $\mathbb{Z}_n$ is enhanced to $U(1)$. The analytic continuation required to compute $\see$, from integer $n$ to $n = 1 + \delta$ with $\delta \ll1$, is more delicate with only the $\mathbb{Z}_n$ symmetry.} Eq.~\eqref{eq:eesphere} depends only on $b$ when $D=3$, and on $b$ and $d_2$ when $D \geq 4$.

As another check of eq.~\eqref{eq:eesphere}, we consider the holographic DCFT given by a probe brane along an AdS$_3$ submanifold inside AdS$_{D+1}$ of radius $L$, with action $S_{\textrm{probe}}$ above. In that case, ref.~\cite{Jensen:2015swa} found
\beq
S_{\rm EE}^{\rm def}= \frac{4\pi T_{\rm br}L^3}{D-1} \ln \left(\frac{\ell}{\varepsilon}\right)+\mathcal{O}(\varepsilon^0),
\eeq 
which agrees with eq.~\eqref{eq:eesphere}, using $b$ and $d_2$ from Table~\ref{tab:ccs}.

Eq.~\eqref{eq:eesphere} agrees with a key result of ref.~\cite{Kobayashi:2018lil} that the coefficient of $\ln \left( \frac{\ell}{\varepsilon}\right)$ in $\sdef$ includes a contribution from the stress-tensor one-point function. We have extended that result, showing that this contribution is proportional to the defect central charge $d_2$.

For defects of various dimensions, refs.~\cite{Kumar:2017vjv,Kobayashi:2018lil,Rodgers:2018mvq} found that the defect's universal contribution to EE need not obey a defect c-theorem. For a 2D defect in a $D \geq 4$ CFT, eq.~\eqref{eq:eesphere} shows why: although $b$ obeys a defect c-theorem, the combination of $b$ and $d_2$ in eq.~\eqref{eq:eesphere} need not, and indeed does not, as ref.~\cite{Rodgers:2018mvq}'s examples show.

\textit{Holographic Examples.} Using eqs.~\eqref{eq:T1pt} and~\eqref{eq:eesphere}, we will extract novel results for $b$ and $d_2$ from existing results for $\langle T^{\mu\nu} \rangle$ and $\sdef$ in two holographic examples of 1/2-BPS 2D surface operators.

First is 4D $\mathcal{N}=4$ SUSY $U(N)$ YM theory at large $N$ and large 't Hooft coupling $\lambda$, dual to 10D type IIB supergravity (SUGRA) on AdS$_5 \times \mathbb{S}^5$~\cite{Maldacena:1997re}. SUGRA solutions describing the most general 1/2-BPS 2D surface operators appear in ref.~\cite{Gomis:2007fi}. Generically such a surface operator breaks $U(N)  \to \prod_{k=1}^n U(N_k)$ with $\sum_{k=1}^n N_k = N$, and produces a non-zero expectation value for one adjoint complex scalar field, $\Phi$, which decomposes into the block diagonal form
\begin{equation}
    \langle \Phi \rangle= \frac{e^{-i \phi}}{\sqrt{2} |x^i|} \mathrm{diag} \left(
        z_1 \mathbb{1}_{N_1}, z_2 \mathbb{1}_2,
        \dots,
        z_n \mathbb{1}_{N_n}
    \right),
\end{equation}
with $\mathbb{S}^1$ angular coordinate $\phi$ around the defect, $z_k \in \mathbb{C}$ dimensionless parameters, and $\mathbb{1}_{N_k}\) the \(N_k \times N_k$ identity matrix~\cite{Gukov:2006jk,Gomis:2007fi,Drukker:2008wr,Drukker:2008jm}. For such a defect ref.~\cite{Drukker:2008wr} holographically computed $\langle T^{\mu\nu}\rangle$ for ${\cal M} = \text{AdS}_3 \times \mathbb{S}^1$, though the result is scheme-dependent. We fix the scheme by conformally mapping AdS$_3 \times \mathbb{S}^1$ to Minkowski space and demanding that without a defect $\langle T^{\mu\nu}\rangle = 0$. Ref.~\cite{Gentle:2015ruo} holographically computed $\sdef$ for a sphere centered on the defect. Using these results, eqs.~\eqref{eq:T1pt} and~\eqref{eq:eesphere} give
\begin{align}
\label{E:4dcharges}
    b &= 3 \left( N^2 - \sum_{k=1}^n N_k^2 \right),
    \\
    \nn
    d_2 &= 3 \left( N^2 - \sum_{k=1}^n N_k^2\right) + \frac{24 \pi^2 N}{\lambda} \sum_{k=1}^n N_k |z_k|^2.
\end{align}
Both of these are manifestly positive, and $b$ is independent of the marginal parameters $\lambda$ and $z_k$.

As discussed in ref.~\cite{Drukker:2008wr} the one-loop $\langle T^{\mu\nu}\rangle$ on AdS$_3\times \mathbb{S}^1$ in the presence of the surface operator matches the term $\propto N/\lambda$ in eq.~\eqref{E:4dcharges}. Given that the other terms in eq.~\eqref{E:4dcharges} are independent of the marginal parameters at large $\lambda$, and that $T^{\mu\nu}$ on AdS$_3\times \mathbb{S}^1$ is scheme-dependent, $b$ and $d_2$ may in fact be one-loop exact. 

Our second example is 1/2-BPS surface operators in the 6D $\mathcal{N}=(2,0)$ SUSY CFT with gauge algebra $\mathfrak{su}(M)$, arising on the worldvolume of $M$ coincident M5-branes. We consider so-called Wilson surface defects~\cite{Ganor:1996nf}, generalizations of Wilson lines to the 6D $\N=(2,0)$ SUSY CFT, which are specified by an $\mathfrak{su}(M)$ representation and a 2D surface. When $M\gg1$ the theory is holographically dual to 11D SUGRA on AdS$_7 \times \mathbb{S}^4$~\cite{Maldacena:1997re}, and Wilson surfaces are dual to M2-branes, or M5-branes with M2-brane flux, reaching the AdS$_7$ boundary at the 2D surface~\cite{Lunin:2007ab,Chen:2007ir,DHoker:2008rje,Bachas:2013vza,Mori:2014tca}.

Ref.~\cite{Gentle:2015jma} holographically computed $\langle T^{\mu\nu} \rangle$ in the presence of a flat Wilson surface, and refs.~\cite{Gentle:2015jma,Estes:2018tnu} holographically computed $\sdef$ for a sphere centered on a flat Wilson surface. Using these results, eqs.~\eqref{eq:T1pt} and~\eqref{eq:eesphere} give
\beq
\label{eq:wilson_surface_b}
b = 24 (w,\varrho) + 3 (w,w), \quad d_2 = 24 (w,\varrho)+ 6 (w,w),
\eeq
where $w$ is the $\mathfrak{su}(M)$ representation's highest weight, $\varrho$ is the $\mathfrak{su}(M)$ Weyl vector, and the scalar product $(\cdot,\,\cdot)$ is with respect to the weight space Killing form. Both $b$ and $d_2$ are $\geq 0$ for all $\mathfrak{su}(M)$ representations, and are invariant under the action of the Weyl group, including complex conjugation of the representation. In the defect RG flows triggered by the expectation value of a marginal Wilson surface operator studied holographically in refs.~\cite{Camino:2001at,Gomis:1999xs,Rodgers:2018mvq}, each of $b$ and $d_2$ is larger in the UV than in the IR, consistent with $b$'s c-theorem. However, the linear combination of $b$ and $d_2$ in eq.~\eqref{eq:eesphere} can be larger in the IR~\cite{Rodgers:2018mvq}, as mentioned above.

\textit{Thermal Entropy.} For a 2D CFT on $\mathbb{S}^1$ of radius $r$, Cardy showed that in the thermodynamic limit $rT \to \infty$, $c$ determines the thermal entropy: $S = \frac{\pi}{6} \, c \, T (2 \pi r)$~\cite{Cardy:1984ql, Affleck:1986zt}. Do $b$, $d_1$, and $d_2$ similarly determine a 2D boundary or defect's contribution to $S$?

Consider the 3D BCFTs of free, massless real scalar or Dirac fermion on a hemi-sphere of radius $r$. In Appendix B we calculate the boundary contribution to thermal entropy, $S_{\partial}$. In the limit $r T \to \infty$ we find
\begin{align}\label{eq:scalar-s-body}
S^{\textrm{R/D}}_{\partial}= \pm \frac{\pi}{12} \, T \, (2 \pi r)\,, \qquad S^{\textrm{f}}_{\partial} = 0
\end{align}
where the superscripts denote the Robin scalar, Dirichlet scalar, and Dirac fermion, respectively. Table~\ref{tab:ccs} shows the Dirac fermion has $d_1\neq0$, so $S^{\textrm{f}}_{\partial}=0$ proves that $S_{\partial}$ cannot have a term $\propto d_1$ with universal non-zero coefficient. Instead, table~\ref{tab:ccs} and eq.~\eqref{eq:scalar-s-body} suggest $S_{\partial} \stackrel{?}{=} \frac{4\pi}{3}b\,T \,(2 \pi r)$, which, if true, looks like $8$ times a Cardy entropy.

However, consider the holographic DCFT given by a probe brane along an asymptotically $AdS_3$ submanifold inside an $AdS_{D+1}$-Schwarzschild black hole of radius $L$ and temperature $T$, with action $S_{\textrm{probe}}$ above. In Appendix C we compute this defect's contribution to $S$,
\beq
S_{\textrm{def}} = \frac{16\pi^2}{D^2} \, L^3 \, T_{\rm br}  T \, (2 \pi r),
\eeq
which via table~\ref{tab:ccs} we can write as $S_{\textrm{def}} \stackrel{?}{=} \frac{1}{D^2}\frac{8\pi}{3} \, b \, T (2\pi r)$, although the probe brane has $b=d_1=d_2$~\cite{Graham:1999pm}, so without further input this choice is arbitrary. We can compare to a DCFT given by gluing two free-field 3D BCFTs along their boundaries, with no boundary interactions, whose $S_{\textrm{def}}$ is simply a sum of the $S_{\partial}$ in eq.~\eqref{eq:scalar-s-body}. Crucially, when $D=3$, no such sum can produce the $1/D^2 = 1/9$ factor in the holographic $S_{\textrm{def}}$. This proves that if $S_{\textrm{def}} \propto b T (2\pi r)$, then the coefficient cannot be universal.


\textit{Acknowledgements.} We are pleased to thank M.~Buican, A.~Chalabi, S.~Dowker, N.~Drukker, T.~Hartman, T.~Faulkner, and J.~Sisti for helpful discussions. We especially thank J.~Estes, C.~Herzog, D.~Krym, and M.~Gutperle for comments on the draft. K. J. is supported in part by the Department of Energy under grant number DE-SC0013682. A.~O'B. is a Royal Society University Research Fellow. B.~R. and R.~R. acknowledge support from STFC through Consolidated Grant ST/L000296/1


\appendix

\textit{Appendix A: Parity-odd central charges.} If we define parity as reflection of the spatial coordinate along $\Sigma$, then WZ consistency allows parity-odd terms in the Weyl anomaly only in $D=4$,
\beq
\label{eq:podd}
\langle T^{\mu}_{~\mu} \rangle_{\textrm{def}} \supset  \frac{\epsilon^{ab} }{2\pi}\left( \tilde{b} \, {\cal R}^N_{ab} + \tilde{d}_1 \epsilon_{ij} \mathring{\Pi}^i{}_{ac} \mathring{\Pi}^j{}_b{}^c + \tilde{d}_2 \epsilon^{ij} W_{aibj}\right),
\eeq
where $\epsilon^{ab}$ and $\epsilon^{ij}$ are the epsilon tensors on $\Sigma$ and the transverse space, respectively, ${\cal R}^N_{ab}$ is the Abelian normal bundle curvature, and $\tilde{b}$, $\tilde{d}_1$, and $\tilde{d}_2$ are parity-odd defect central charges. More precisely, $\tilde{b}$, $\tilde{d}_1$, and $\tilde{d}_2$ are odd under parity and time reversal, and even under charge conjugation. All three terms in eq.~\eqref{eq:podd} are type B in the classification of ref.~\cite{Deser:1993yx}. The integral of the first term in eq.~\eqref{eq:podd} is the Euler class of the normal bundle, which is a topological invariant. For our DCFTs, $SO(2,2)\times SO(D-2)$ symmetry forbids any parity-odd tensor contributions to $\langle T^{\mu\nu}\rangle$ or to ${\cal D}^{\mu}$'s 2-point function, so $\tilde{b}$, $\tilde{d}_1$, and $\tilde{d}_2$ do not appear in these correlators.  We have written eq.~\eqref{eq:podd} in a form reminiscent of the parity-even anomaly in eq.~\eqref{eq:defect-A}, but in fact two of the terms in eq.~\eqref{eq:podd} are not independent. Contracting the Ricci equation with $\epsilon^{ab} \epsilon^{ij}$ gives $\epsilon^{ab} ( {\cal R}^N_{ab} - 2  \epsilon_{ij} \mathring{\Pi}^i{}_{ac} \mathring{\Pi}^j{}_b{}^c  + \epsilon^{ij} W_{ai bj} ) = 0\)~\footnote{We thank A.~Chalabi for pointing this out.}. As a result, one of the terms in eq.~\eqref{eq:podd} may be replaced by a linear combination of the other two terms, so that only two of the parity-odd defect central charges are independent.

We have not yet found any examples in which $\tilde{b}$, $\tilde{d}_1$, and $\tilde{d}_2$ are non-trivial. One candidate is 4D gauge theory CFTs with gauge field BCs that produce a 2D defect equivalent to a 2D theta angle~\cite{Gukov:2006jk}. However, these defects are in fact even under our parity transformation, so all parity-odd defect central charges vanish.

Another candidate is 10D type IIB SUGRA on $AdS_5 \times S^5$ with a probe D7-brane along AdS$_3 \times \mathbb{S}^5$. The dual DCFT is 4D $\mathcal{N}=4$ $U(N)$ SUSY YM theory with 1/2-BPS coupling to a 2D $\N=(0,8)$ multiplet, whose only on-shell DOF is a chiral fermion~\cite{Harvey:2007ab,Buchbinder:2007ar,Harvey:2008zz}. When the D7-brane's worldvolume gauge field vanishes, the parity-odd terms in the D7-brane's action are WZ terms of the schematic form $\int P[C_4] \wedge \left[p_1({\cal R}^T) - p_1({\cal{ R}}^N) \right]$, with Ramond-Ramond 4-form $C_4$ and tangent and normal bundle curvatures ${\cal R}^T$ and ${\cal R}^N$, respectively. Compactification on the $\mathbb{S}^5$ produces an AdS$_3$ probe brane with gravitational Chern-Simons terms for the induced metric and for ${\cal{R}}^N$, which by inflow correspond to gravitational and normal bundle anomalies for the dual defect, respectively. However, a straightforward holographic calculation~\cite{Graham:1999pm} shows that these anomalies do not contribute to the Weyl anomaly, so again in this example all parity-odd defect central charges vanish.

We leave the physical significance of these parity-odd defect central charges---if any---for future research.

\textit{Appendix B: Free Field Thermal Entropy.} 
In this appendix we compute the thermal entropy $S$ of a 3D conformally coupled free, massless, real scalar field or Dirac fermion on a hemisphere of radius $r$, $\mathbb{HS}^2_r$. More precisely, we compute the boundary contribution $S_{\partial}$ to $S$.

Let $\chi$ be a free, massless, real scalar. We will use the imaginary time formalism, meaning we work in Euclidean signature. We consider $\chi$ on the manifold ${\cal M} = \mathbb{S}^1_{\beta} \times \mathbb{HS}_r^2$, where $\mathbb{S}^1_{\beta}$ is the periodic Euclidean time direction $\tau \sim \tau + \beta$ with inverse temperature $\beta = 1/T$, with metric
\begin{align}
ds^2 = d\tau^2 + r^2(d\theta^2+\sin^2\theta d\phi^2),
\end{align}
where $\theta\in[0,\,\frac{\pi}{2}]$ with boundary at $\theta = \pi/2$, and $\phi\in[0,\,2\pi)$. The scalar's action is
\begin{align}
S_{\chi}= \int_{{\cal M}} d^3x \sqrt{g}\left(\frac{1}{2}(\partial\chi)^2 +\frac{\cal{R}}{8}\chi^2\right) + \int_{\partial} d^2y\sqrt{\gamma} ~\frac{\cal{K}}{4}\chi^2,
\end{align}
where $\cal{R}$ is the Ricci scalar on $\cal{M}$ and $\cal{K}$ is the trace of the extrinsic curvature of $\partial\cal{M}\hookrightarrow \cal{M}$. The free energy is
\begin{align}\label{eq:scalar-F-1}
F = -\ln Z =\frac{1}{2}\ln\det{\!}' ~\Box_{\cal{M}}.
\end{align}
We need the eigenvalues of the kinetic operator $\Box_{\cal{M}}$ subject to chosen BCs. Expanding on a basis of Matsubara modes and $\mathbb{S}^2_r$ spherical harmonics, $\chi = \sum_{n,l,m} \chi_{nlm} $, the eigenvalues of $\Box_{\cal{M}}$ are given by
\begin{align}\label{eq:scalar-eigen}
\Box_{\cal{M}} \chi_{n l m} = -\left(\frac{4\pi^2n^2}{\beta^2}+\frac{(l+\frac{1}{2})^2}{r^2}\right)\chi_{n\ell m},
\end{align}
with degeneracy $d_{l}$ determined by BCs. The conformal BCs are Dirichlet (D), $\chi|_{\partial{\cal{M}}}=0$, and Robin (R), $\partial_\theta \chi|_{\partial{\cal{M}}} = {\cal{K}}\chi/4$. If we define a $\mathbb{Z}_2$ operator $P_\theta: \theta \to \pi -\theta$, then $P_\theta  Y_{lm} = (-1)^{l+m} Y_{lm}$. Imposing D or R projects the eigenmodes $\chi_{nl m}$ onto odd or even subspaces of $P_{\theta}$, respectively, with degeneracies $d_l = l$ for D and $d_l = l+1$ for R.  

Inserting eq.~\eqref{eq:scalar-eigen} into eq.~\eqref{eq:scalar-F-1} and using the Poisson re-summation formula
\begin{align}\nonumber
\sum_{n\in\mathbb{Z}} \ln\left( n^2 p^{-2} +q^2\right) = 2\ln \left[2\sinh\left(\pi p |q|\right)\right],
\end{align} 
we can express the free energy as
\begin{align}
F = \sum_{l}d_l\left(\ln\left( 1- e^{-\frac{\beta(2l+1)}{2r}}\right) +\frac{\beta(2l+1)}{4r}\right).
\end{align}
We can then rewrite $F$ as a multiparticle partition function~\cite{Giombi:2014yra},
\begin{align}
F =- \sum_{j=1}^{\infty}\sum_l \frac{d_l}{j} e^{-\frac{j\beta}{2r}(2l+1)}+\sum_\ell d_\ell \frac{(2l+1)\beta}{4r}.
\end{align}
Inserting the degeneracies for D or R and performing the sums over $l$ leads to 
\begin{align}
\begin{split}
F^{\textrm{D}}& = - \sum_{j=1}^{\infty} \frac{e^{j\beta/2r}}{j(1-e^{j\beta/r})^2}- \frac{\beta}{48r},\\
F^{\textrm{R}} &= -\sum_{j=1}^{\infty}\frac{e^{3j\beta/2r}}{j(1-e^{j\beta/r})^2} +\frac{\beta}{16r}.
\end{split}
\end{align} 
Expanding in $\beta/r \ll 1$ (the thermodynamic limit) and retaining only the leading boundary extensive term gives
\begin{align} \label{eq:free_scalar_subtraction}
F^{\textrm{D/R}}_{\partial}= F^{\textrm{D/R}} - \frac{1}{2}F_{\textrm{CFT}}= \mp\frac{\pi}{24} \, T \, (2 \pi r),
\end{align}
where the upper/lower sign corresponds to D/R respectively. As a result,
\begin{align}
\label{eq:scalarS}
S^{\textrm{D/R}}_{\partial} = -\frac{\partial}{\partial T} \left(T F^{\textrm{D/R}}_{\partial}\right)= \mp \frac{\pi }{12} T (2\pi r).
\end{align}

Let $\Psi$ be a free, massless Dirac fermion on ${\cal{M}}$. The calculation of free energy is analogous to that for $\chi$ above, but now with  $F^{\textrm{f}} = - \frac{1}{2}\ln \det \hat{\Box}_\mathcal{M}$, with $\hat{\Box}_\mathcal{M}$ the square of the Dirac operator on \(\mathcal{M}\). We expand $\psi^{\pm} = \sum_{n,l m}\psi^{\pm}_{nl m}$ on a basis of Matsubara modes labelled by \(n\) and spin-weighted spherical harmonics $Y_{s;l m}$. The BCs are more subtle, and the degeneracies of spin-weighted spherical harmonics are different, see e.g. \cite{vanNieuwenhuizen:2012zk}.  Decomposing $\psi = (\psi^+,\,\psi^-)$ into Weyl spinors, in order not to over-constrain the bulk Dirac equation, mixed D and R BCs are needed, e.g. D on $\psi^+$ and R on $\psi^-$~\cite{Csaki:2003sh}. On $\mathbb{S}^1_{\beta} \times \mathbb{S}^2_r$, the degeneracies for $\psi^{\pm}_{nl m}$ are $d_l= 2(l+1)$. Because $P_{\theta} Y_{s;l m} = (-1)^{l+m}Y_{-s; l m}$, imposing BCs evenly lifts the degeneracies to $d_l = l+1$. As a result, the free energy on $\mathbb{S}^1_{\beta} \times \mathbb{HS}^2_r$ is precisely half of that on $\mathbb{S}^1_{\beta} \times \mathbb{S}^2_r$, so the subtraction in eq.~\eqref{eq:free_scalar_subtraction} gives $F^{\textrm{f}}_{\partial}=0$. The boundary therefore makes no contribution to the thermal entropy,
\begin{equation}
    S^{\textrm{f}}_{\partial}= 0.
\end{equation}


\textit{Appendix C: Probe Brane Thermal Entropy.} We consider a CFT on $\mathbb{S}^1_{\beta} \times \mathbb{S}^{D-1}$, where $\mathbb{S}_1$ is a circle of length $\beta$, holographically dual to Einstein gravity in global $AdS_{D+1}$-Schwarzschild black hole, with metric
\begin{align}
\nn
G &= \frac{d\rho^2}{f(\rho)} + f(\rho) d\tau^2 + \rho^2 \left(d\theta^2 + \sin^2 \theta \, d s_{\mathbb{S}^{D-2}}^2 \right), \\
f(\rho) &= 1 + \frac{\rho^2}{L^2} - (\rho_{\textrm{H}}^{D-2} + \rho_{\textrm{H}}^D L^{-2} )\rho^{2-D},
\end{align}
with $\rho\in[\rho_{\textrm{H}},\infty)$ with horizon $\rho_{\textrm{H}}$ at the largest real zero of $f(\rho)$ and $AdS_{D+1}$ boundary at $\rho \to \infty$, $\tau \sim \tau + \beta$, $\theta \in[0,\pi]$, and $L$ the $AdS_{D+1}$ radius. We choose a defining function $L^2/\rho^2$, so the dual CFT lives on a spatial $\mathbb{S}^{D-1}$ of radius $L$. The temperature $T = 1/\beta$ is~\cite{Witten:1998zw},
\beq
\label{eq:horizon_temperature}
T = \frac{D \rho_{\textrm{H}}^2 + (D-2) L^2}{4 \pi \rho_{\textrm{H}} L^2}.
\eeq

We introduce a 2D conformal defect dual to a probe brane along an asymptotically $AdS_3$ submanifold, with Euclidean action
\beq
\label{eq:probe_action}
S_{\textrm{probe}} = T_{\rm br} \int d^3 \xi \sqrt{\textrm{det}P[G_{MN}])} - \frac{T_{\rm br} L}{2} \int d^2 \sigma \sqrt{\det \gamma}\,.
\eeq
with tension $T_{\rm br}$, brane worldvolume coordinates $\xi$, and the second integral is over the intersection of the worldvolume with the $AdS_{D+1}$ boundary, with coordinates $\sigma$ and induced metric $\gamma$. The second integral is required for holographic renormalization.

The equations of motion arising from $S_{\textrm{probe}}$ are solved by a brane which spans the directions $\rho$, $\tau$, and $\theta$, holographically dual to a defect wrapping ${\mathbb{S}}_\beta^1$ and a maximal $\mathbb{S}^{1}\in\mathbb{S}^{D-1}$.  At leading order in the probe limit, the defect's contribution to the free energy is given by $S_{\textrm{probe}}$ evaluated on this solution~\cite{Witten:1998zw},
\beq
F_{\textrm{def}} = - \frac{\pi T_{\rm br}}{T} \left(\rho_{\textrm{H}}^2 + \frac{L^2}{2}\right).
\eeq
The defect's contribution to the thermal entropy is thus
\beq
S _{\textrm{def}}= - \frac{\partial}{\partial T} \left( T F_{\textrm{def}}\right)= \frac{16 \pi^2}{D^2} L^3\,T_{\rm br} \,  T (2 \pi r),
\eeq
where we identified the $\mathbb{S}^1 \in \mathbb{S}^{D-1}$ radius as $L=r$.
 
\bibliography{dcft_v3}

\end{document}